# Anchor Sponsor Firms in Open Source Software Ecosystems

Brigitta Németh[1,2] * & Johannes Wachs [3,2,4]

1. Doctoral School of Economics, Business and Informatics, Corvinus University of Budapest
2. Institute of Economics, HUN-REN Centre for Economic and Regional Studies, Budapest
3. Department of Network Science, Corvinus University of Budapest
4. Complexity Science Hub, Vienna
* Correspondence to: johannes.wachs@uni-corvinus.hu

Firms are intensifying their involvement with open source software (OSS), going beyond contributing to individual projects and releasing their own core technologies as OSS. These technologies, from web frameworks to programming languages, are the foundations of large and growing ecosystems. Yet we know little about how these anchor sponsors shape the behavior of OSS contributors. We examine Mozilla Corporation's role as incubator and anchor sponsor in the Rust programming language ecosystem, leveraging data on nearly 30,000 developers and 40,000 OSS projects from 2015 to 2022. When Mozilla abruptly exited Rust in August 2020, event-study models estimate a negative impact on ecosystem activity: a 9% immediate drop in weekly commits and a 0.6 percentage point decline in trend. We observe an asymmetry in the shock's effects: former Mozilla developers and close collaborators continued contributing relatively quickly, whereas more distant developers showed reduced or ceased activity even six months later. An agent-based model of an OSS ecosystem with an anchor sponsor replicates these patterns. We also find a marked slowdown in new developers and projects entering Rust post-shock. Our results suggest that Mozilla served as a critical signal of Rust's quality and stability. Once withdrawn, newcomers and less-embedded developers were the most discouraged, raising concerns about long-term ecosystem sustainability.

*We thank Zoltán Elekes, Dima Yankova, Balázs Lengyel, participants of the First Danube Workshop on Software and the Digital Economy, the 70th Anniversary Conference of the Hungarian Centre for Economics, and the "Open Source in the Global Digital Economy" 2024 OFA Symposium for helpful suggestions and comments. The authors acknowledge support from the Hungarian National Scientific Fund (OTKA FK-145960) and from the Mentorship Program of the HUN-REN CERS Institute of Economics.*

Firms engage so enthusiastically with open source software (OSS) in part because, as Bill Joy—co-founder of Sun Microsystems—once quipped, "no matter who you are, most of the smartest people work for someone else" (Lakhani and Panetta, 2016). By reaching beyond their own boundaries firms expand their capacity for innovation and tap into a broader range of ideas (West and Bogers, 2014). And it works: research shows that companies—whether startups or established firms—that use and contribute to OSS tend to be more productive and competitive (Conti et al., 2023; Nagle, 2019, 2018; Wright et al., 2023). Given that OSS projects are valuable public goods (Gortmaker, 2024; Greenstein and Nagle, 2014) requiring constant upkeep (Eghbal, 2016), the growing involvement of firms in OSS is an important trend in the digital economy.

However, firm involvement in OSS is also changing in its depth. Increasingly firms spin out key technologies like web frameworks, deep learning libraries, and even entire programming languages as OSS. These projects become the foundations for entire OSS ecosystems which attract vibrant communities of users and contributors. For example, Meta supports projects such as Llama, React, and PyTorch; Google supports Tensorflow and Kubernetes; and Microsoft supports Visual Studio Code and TypeScript. Often the software that others build on these core technologies goes well beyond the scale and scope of the initial project. We refer to these individual firms which incubate and launch specific OSS ecosystems as *anchor sponsors*.

But how does this trend of anchor sponsorship influence the behavior of other contributors—those smart people who work for someone else? Indeed, firm participation is known to affect how OSS is made (Li et al., 2024), changing developer incentives and the structure of communities. Firm involvement may discourage volunteers from contributing to a project (Birkinbine, 2020) and crowd-out other contributors, reducing project sustainability (Zhang et al., 2022b). Indeed most research on OSS frames it as a decentralized peer production community of contributors whose intrinsic motivations tend a creative flame (Benkler, 2002; Bonaccorsi and Rossi, 2004; Lakhani and Wolf, 2005; Raymond, 1999). In other words, the collective and networked efforts of individuals (Dabbish et al., 2012) and their diverse motivations (Gerosa et al., 2021), ranging from signaling (El-Komboz and Goldbeck, 2024; Lerner and Tirole, 2002; Riehle, 2015) to "scratching an itch" (Raymond, 1999), make OSS successful. A recent study of the GitHub Sponsors program, which enables OSS developers to collect funds from the crowd (Conti et al., 2023), nicely highlights the complexity of motivations and incentives in OSS: once crowdfunded, developers tend to focus on existing projects and are less likely to start new ones.

On the other hand, these results focus on firm involvement in specific projects or on developer motivations to contribute to OSS in general, and implications for how anchor sponsors influence other developers remain unclear. In the anchor sponsor context, developers are often choosing between ecosystems in which to realize a project. So we must consider why developers use or contribute to a specific project or ecosystem as opposed to alternatives. Here the literature is less developed, though it is clear that the size and quality of the community is important (Ma et al., 2020). We know that developers carefully consider both social and technical aspects of a project when considering an investment of their time and effort (Dabbish et al., 2012). Indeed, because developers mostly write code that builds on and depends on other code in an ecosystem (Decan et al., 2019), developers search for signals that an ecosystem they are considering joining will remain stable or grow. New ecosystems must offer significant advantages to outweigh the stability established systems provide. This raises the question of how a prominent anchor sponsor influences developer engagement and shapes an ecosystem's evolution.

To explore this question, we draw an analogy to the notion of an "anchor firm" from the regional studies literature[1] (Agrawal and Cockburn, 2003; Feldman, 2003). Anchor firms are large actors in a local region or cluster which create important externalities that influence the behavior of other firms (Nilsson et al., 2024). Anchor firms are thought to energize clusters by creating scale effects, knowledge spillovers (Gong et al., 2024), and by signaling quality and suggesting stability (Crescenzi et al., 2022). They are also thought to shape the evolution of clusters substantially.

Many of these externalities have analogies in the OSS ecosystem context. The software industry in general is a classic example of a sector with increasing returns to scale (Arthur, 1994). In other words, the more people using and contributing to an ecosystem, the more useful and attractive it becomes. Commercial sponsors provide an initial boost to scale (Riehle, 2007). OSS developers are also highly motivated by the chance to learn from others (Gerosa et al., 2021), suggesting how knowledge spillovers out of the anchor sponsor can attract new contributors (Greenstone et al., 2010). Finally, the anchor sponsor can signal quality and stability: if a reputable tech firm is investing heavily in a specific ecosystem, it suggests that the ecosystem will not disappear overnight.

Against this theoretical backdrop, we carry out an empirical study of an OSS ecosystem with an anchor sponsor, specifically the Rust programming language. Rust was incubated in the early 2010s by the Mozilla Corporation, a large tech company best known for developing the Firefox web browser. Following its mainstream release in 2015, Rust quickly grew into a popular and widely used programming language, with tens of thousands of developers contributing to a growing ecosystem of OSS projects using the language. Critically, we exploit the sudden exit of Mozilla in August 2020 (Mozilla Corporation, 2020) to study changes in developer behavior in the presence and absence of an anchor sponsor.

Specifically, we analyze the impact of Mozilla's layoffs on the Rust ecosystems using a comprehensive longitudinal data on contributions to OSS libraries in Rust[2] (Schueller et al.,

---

[1] The regional studies literature itself borrowed the term from the real estate literature on "anchor tenants" - large stores in shopping malls which drive foot traffic and facilitate the entry of niche stores.

[2] Our dataset covers the vast majority of libraries hosted on Cargo, the primary package manager of the Rust programming language.

2022). In Figure 1 we visualize the smoothed time series of commits, elementary code contributions in OSS, to all OSS Rust libraries from the language's public 1.0 release in 2015 up until 2022. We highlight the Mozilla layoffs and observe a striking break in the trend of activity from quick growth to something more gradual. Our main statistical analysis confirms the significance of this change in trend. We also observe significant heterogeneities. Developers formerly employed by Mozilla, as well as those closely connected to them in the collaboration network, quickly rejoined or continued contributing to the ecosystem. In contrast, activity declined among developers farther removed from Mozilla affiliates. Additionally, we find a slower influx of new users and projects into the Rust ecosystem following the withdrawal of its anchor sponsor. This suggests a strong commitment among the most embedded contributors but indicates potential challenges in attracting new participants, which is essential for the ecosystem's growth, sustainability, and ability to innovate.

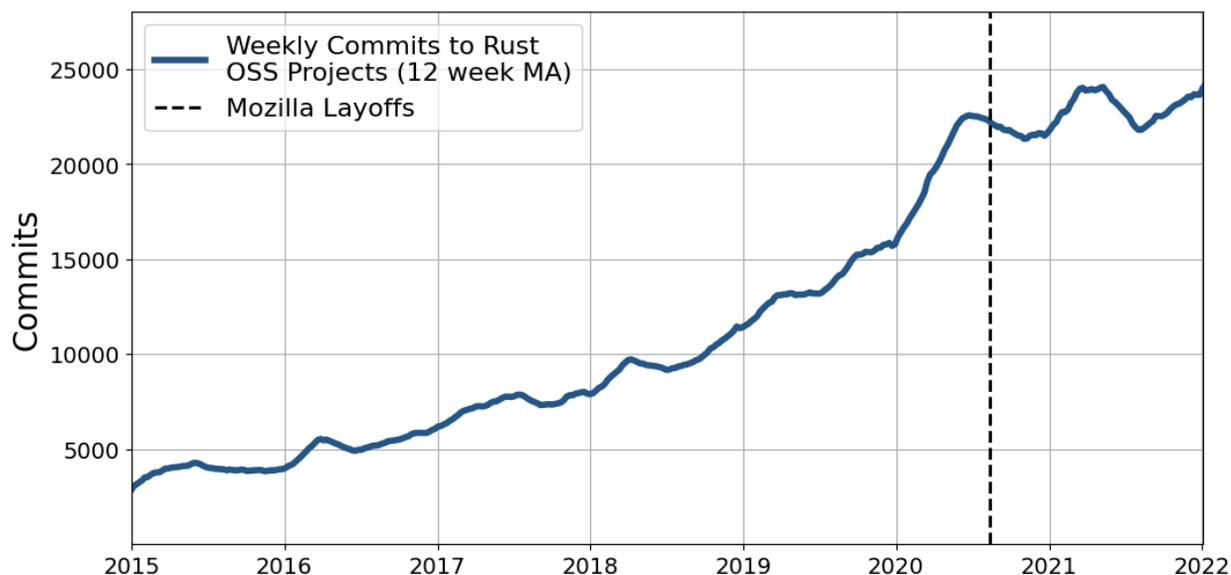

*Figure 1: The evolution of the Rust ecosystem 2015-2022. We plot the 12 week moving average of the number of commits, elementary code contributions, made by contributors to OSS packages in Rust. The layoffs of a large share of Mozilla's Rust team in August 2020 is highlighted.*

The rest of the paper is organized as follows. We first review related work on firm involvement in OSS. We then describe Mozilla's involvement in the Rust ecosystem and introduce research questions. We then describe the data and methods, and report results. We present an agent-based model of developer contributions to an OSS ecosystem with an anchor sponsor, replicating our empirical findings qualitatively. We conclude with a discussion of implications for firm involvement in OSS.

# Firm involvement in OSS

While software is generally known to require maintenance to preserve its functionality over time (Lehman, 1984), in open source ecosystems human resources cannot easily be allocated to fix specific problems as there are no direct market mechanisms or formal hierarchies in play. The research literature on maintenance of OSS projects tends to focus on key individuals (Miller et al., 2019), for example defining a project's Truck Factor (Avelino et al., 2016; Williams and Kessler, 2003), or on technical dependencies (Decan et al., 2019), rather than on the organizational models that may influence OSS sustainability (Riehle, 2007).

Recently, a variety of such models have emerged to support OSS development including crowdfunding through GitHub Sponsors (Conti et al., 2023) and support from the public sector like the Open Technology Fund in the USA or the Next Generation Initiative in EU (Osborne, 2024). Private funding, provided by companies and philanthropic foundations through grants, bounty programs, sponsorships, or membership in foundations, is significant due to its prevalence and impact. There is a wide range of what can be considered involvement: firms may sponsor or hire individuals to work on specific projects, or pursue large-scale involvement of a firm's developers in specific projects and ecosystems (Barcomb et al., 2018).

But why do firms engage in OSS? The use of open source software has a positive impact on value-added productivity in firms that are IT-intensive or operate in IT-producing industries (Nagle, 2019). By actively contributing, firms gain significant productivity benefits through learning by doing, giving them a competitive advantage over free-riders. Involvement in OSS allows companies greater control and flexibility to tailor software to their unique needs, enhancing adaptability in fast-changing markets. Contributing also helps firms build valuable relationships and reputational capital within the community (Nagle, 2018), which can also facilitate labor market matching. OSS contributions are also predictive of better outcomes for start-ups like acquisition (Wright et al., 2024) because of the valuable information about a firm's quality and capabilities contained in their open code.

A fundamental challenge for firms is that most of the relevant knowledge needed to solve problems resides outside their organizational boundaries (Lakhani et al., 2007). Distributed innovation systems provide an alternative approach to accessing this external knowledge. OSS communities are the among most fully developed examples of distributed innovation systems, characterized by decentralized problem-solving, self-selected participation, self-organizing coordination and collaboration, and the free revealing of knowledge (Lakhani and Panetta, 2016). However, hybrid organizational models that blend community with commercial success require a fundamental reorientation of views about incentives, task structure, management, and intellectual property (Lakhani and Panetta, 2016). Effective boundary management is crucial for influencing the community to utilize the firm's resources and ensuring that community innovation efforts focus on areas the firm is interested in absorbing, preventing wasted efforts (Teigland et al., 2014).

Firm involvement in core OSS libraries has become significant (Mehler et al., 2024). Many well-known OSS ecosystems such as the Linux kernel, Android, and OpenStack are developed mainly through collaborations of different companies. For example, companies contribute more than 90% of code on average in each version of OpenStack (Zhang et al., 2022a). In a study of the Atom, Electron, Hubot, git-lfs, and linguist GitHub projects, researchers found that employees are responsible for 43% of the pull requests performed. Even with high support from the external community, firm employees play the central roles in these projects (Dias et al., 2018). Companies often join forces to support OSS technologies through foundations, for example the Apache, Eclipse, and Gnome Foundations (Riehle, 2010). This has led to OSS communities evolving from networks of individuals to networks of companies (Ågerfalk and Fitzgerald, 2008), where 'co-opetition'—a mix of collaboration and competition—has become a defining characteristic (Osborne et al., 2024). These strategic alliances, referred to as "access relationships," allow firms to leverage shared resources while maintaining competitive advantages (Stuart, 2000).

## Consequences of firm involvement

Having established that firm involvement in OSS projects is both significant and long-established, we now consider potential consequences. Indeed, OSS ecosystems are thought to thrive on collaborative relationships between participating firms and the broader community, combining structured business strategies with decentralized problem-solving to drive innovation (Fitzgerald, 2006). However, while firms bring valuable resources, their participation also introduces trade-offs that can shape the long-term sustainability of OSS ecosystems.

One key concern is the shift from intrinsic to extrinsic motivations in OSS production. Traditional intrinsic drivers such as intellectual challenges, play value (Hertel et al., 2003; Lakhani and Wolf, 2005), the gift culture surrounding OSS (Bitzer et al., 2007; Zeitlyn, 2003) or the sense of belonging (Trinkenreich et al., 2024) might be overshadowed or crowded out. While monetary rewards can have positive side effects such as aiding in recruiting new employees (Alexy, 2009) and individuals are more likely to expect to enhance their reputation if they write valuable code seen by a large community, including large software firms (Lerner and Tirole, 2002), financial incentives do not always lead to greater innovation. For instance, crowdfunded OSS developers have been found to be less likely to innovate (Conti et al., 2023).

While some studies suggest that company-dominated projects deter new contributors—who may fear providing free labor to the primary sponsor Zhang et al. (2022a)—economic research on clusters and regional economies presents a contrasting perspective which we believe can be useful to understand the role of firms in OSS ecosystems. Specifically the anchor hypothesis suggests that large, established firms generate scale and externalities that benefit smaller players and increase overall innovative output in clusters. This effect was first studied in real estate economics (Bean et al., 1988), and later used to understand R&D activity (Agrawal and Cockburn, 2003), and clusters in the biotech industry (Feldman, 2003).

Applying this concept to OSS, an established anchor sponsor may make an open-source ecosystem more attractive to newcomers by signaling long-term stability, providing access to resources, and fostering an environment where contributors can build valuable skills and reputational capital. The presence of a well-known corporate backer can legitimize a project, encouraging new developers to join with the expectation of sustained activity, mentorship, and potential career opportunities.

## The Risk of Firm Withdrawal

Beyond influencing contributor dynamics, corporate involvement in OSS introduces the risk of withdrawal, leaving widely-used code unmaintained. In OpenStack withdrawals are not uncommon: more than half of the companies that joined in a certain version of OpenStack withdrew later (Zhang et al., 2022a). Withdrawal can affect code quality and sustainability in multiple ways: historical code will be left unmaintained, and productivity in the project will decrease following the departure of experienced developers (Mockus, 2009). In cases of single-company domination, the departure of the company often jeopardizes the survival of the project (Zhang et al., 2018; Zhou and Mockus, 2014).

Unlike individual contributors, whose departures are often gradual, a firm's withdrawal may be more sudden and disruptive. This can impose a front-loaded burden on remaining developers, who are left with a dramatically increased workload, and in some cases, it may lead to the ultimate failure of a major project. Moreover, external developers may hesitate to join a project if they perceive instability in its trajectory (Qiu et al., 2019a; Smirnova et al., 2022), and companies evaluating new OSS participation also consider the stability of the ecosystem (Butler et al., 2022).

Firm involvement in OSS is extensive, with significant contributions to major ecosystems and substantial benefits for both companies and open-source communities. However, corporate participation introduces trade-offs, including shifts in motivation dynamics and the risk of abrupt withdrawal. While firms can serve as anchors that stabilize and legitimize projects, their exit can jeopardize long-term sustainability (Ornston and Camargo, 2024). Addressing these challenges requires further research into firm withdrawal's systemic effects and strategies to ensure resilient OSS ecosystems. This gap is particularly relevant given the importance of OSS to the digital economy and the dependence of sustainability on effective maintenance and a healthy developer community (Eghbal, 2016).

## The Case of Mozilla and Rust

Mozilla, originally a subsidiary of Netscape launched in the late 90s, has a long track record of participation in OSS. Best known for the open-source Firefox web browser, Mozilla became a successful multi-project, community-oriented entity in the OSS world in the 2000s, engaging episodic collaborators by cooperating with non-profit organizations to promote values such as openness and efficacy (Barcomb et al., 2018). Mozilla community managers and infrastructure such as the Mozilla Bootcamp and Hive Learning Networks contributed to Mozilla's image as a responsible and attractive collaborator for OSS developers (Smirnova et al., 2022).

Mozilla invested a significant amount of resources into incubating and developing the Rust programming language, which was started as a personal project in 2006 by a Mozilla employee (Thompson, 2023). Mozilla's management was interested in the potential of Rust's memory-safety features to improve its web browser technology, and allocated several developers to support the project from 2009. By the time of the language's public 1.0 release, Mozilla had a core team of 9 developers working on the project, with 30-40 developers from other teams making contributions (Klabnik, 2016).

Upon its release Rust quickly became a successful programming language (Perkel, 2020). It was voted the "most loved" programming language in several iterations of the Stack Overflow user survey[3], is one of the fastest growing languages on GitHub[4], and is one of three languages that can be used within Linux kernel software (besides assembly and C)[5]. Like many languages, Rust has a primary package manager called Cargo which allows developers to share and interlink their projects. Rust's success as a bonafide OSS ecosystem beyond the scope of Mozilla, is reflected in our data: in the first half of 2020 only 1.1% of all commits to Rust OSS libraries came from 92 Mozilla developers we could identify using email suffix data (i.e. @mozilla.com). On the other hand, those developers made 1.7% of commits to libraries among the top decile downloaded in the same time period, signaling that they were working on the most central and vital libraries in the ecosystem.

Despite this success, changes in the competitive landscape and shifts in user preferences had put pressure on Mozilla's revenue streams, leading to a restructuring resulting in the layoff of many developers working on Rust on August 11, 2020. This case provides a valuable opportunity to study how centralized sponsorship of an OSS ecosystem influences the behavior of contributors. As a trusted and established participant in OSS, Mozilla's brand certainly lent credibility to the Rust ecosystem.

---

[3] https://survey.stackoverflow.co/2020#technology-most-loved-dreaded-and-wanted-languages
[4] https://github.blog/developer-skills/programming-languages-and-frameworks/why-rust-is-the-most-admired-language-among-developers/
[5] https://rust-for-linux.com/

## Research questions

While we expect and indeed do observe a significant drop in contributions from contributors using @mozilla.com email addresses (see Figure 2 below), we want to observe the effect of the layoffs on the broader ecosystem. At the macro-level we want to know if activity decreased relative to a counterfactual of no lay-offs. Figure 1 offers suggestive evidence of a decrease in the growth rate of contributions after August 2020. We also expect to see heterogeneous impacts on ongoing activity versus new activity (i.e. the number of first time contributors and new projects created).

At the user level, we study changes in individual behavior as a function of their proximity to the Mozilla Rust development team. We quantify this using distance in the developer collaboration network and interpret it as a measure of embeddedness in the core of the Rust ecosystem. Contributors closer to (former-)Mozilla developers are highly active and invested in the ecosystem: in the period before the shock, direct collaborators of Mozilla developers contributed on average 211 commits (median: 42), compared with 137 commits on average for two-step neighbors (median: 28) and 90 for three-step neighbors (median: 23). Are they more or less likely to decrease their contribution activities? On the one hand, such users have more invested into Rust and have greater social capital connected to its ecosystem and thus may be more likely to stay (Qiu et al., 2019b). On the other hand, they may be highly skilled and more able to switch between languages, or be directly impacted by the sudden obsolescence of specific projects.

Additionally, we distinguish between working-hour and non-working-hour developers to capture differences in participation based on time commitment and potential motivations (Riehle et al., 2014). "Working-hour" contributors often exhibit patterns consistent with paid or semi-professional involvement, while "non-working-hour" contributors are more likely to be volunteers whose participation is driven by intrinsic motivations. This categorization helps us assess whether corporate sponsorship influences OSS activity primarily through signaling credibility to other firms or if it also substantially motivates volunteer contributions. Together, these groupings enable us to comprehensively assess Mozilla's impact across distinct user segments, shedding light on which aspects of the community were most vulnerable to the withdrawal of institutional support.

## Data and Methods

We use a nearly comprehensive database of activity in the Rust OSS ecosystem (Schueller et al., 2022). The database uses Cargo, the primary package manager of Rust, to define a universe of OSS packages. The database includes data on all commits (the elementary code contributions) to these packages, harvested from their repositories on platforms including GitHub, Gitlab, and BitBucket. By integrating diverse sources, we avoid biases introduced by solely using GitHub (Trujillo et al., 2022).

Data harvested from these traces enable the study of large-scale decentralized collaborative work and the impact of the shock on the Rust ecosystem as a whole. Crucially, we can identify likely current and former Mozilla employees by the email suffixes on the contributions they make. Email prefixes and GitHub/Gitlab user ids are anonymized to preserve user privacy. In Figure 2 we plot the share of commits signed with @mozilla.com email addresses, highlighting the timing of the Mozilla layoffs.

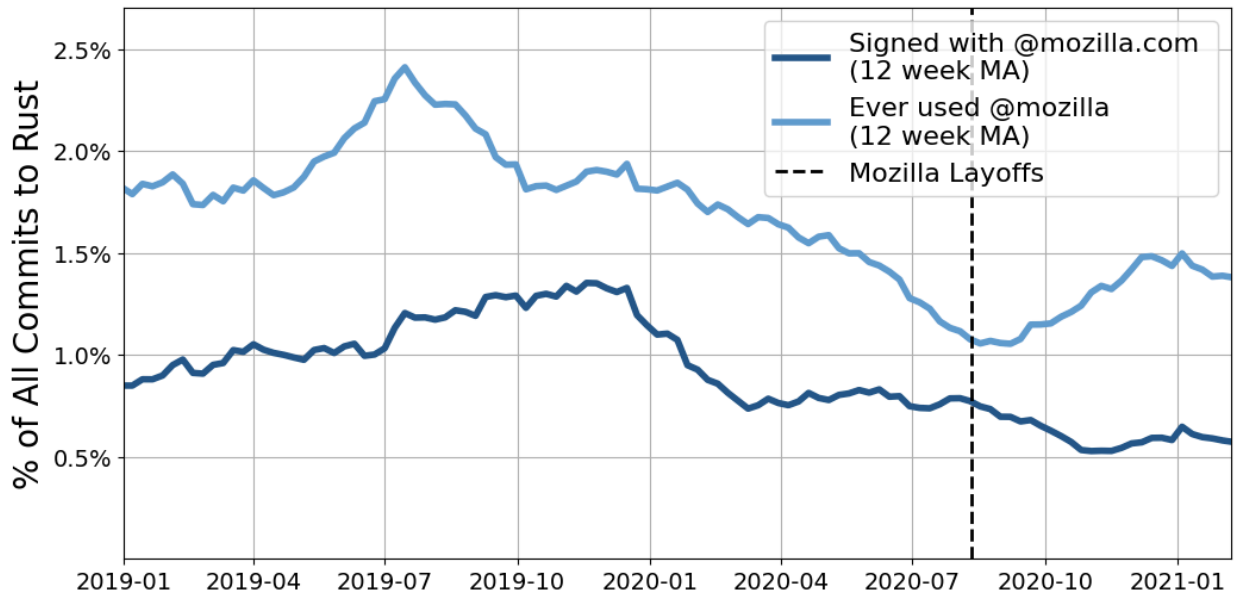

*Figure 2: Share of commits (contributions) to the Rust OSS ecosystem signed with @mozilla.com email addresses, and the share of commits from user accounts who ever sign a commit with such an address. Persistent user accounts sign their commits with varying email addresses.*

Crucially, our analysis proceeds at the user, rather than email address level, as users may make contributions from multiple email addresses. For example, a user may have a GitHub account linked to a work email, which he or she changes when switching jobs. This allows us to track different kinds of users over time in a consistent manner. To model the consequences of the Mozilla shock, we focused on the period between January 1, 2019. and February 8, 2021. The dataset includes 24,245 libraries to which 25,413 developers made 1,977,887 commits. The data is curated to exclude bot accounts and their contributions, and to disambiguate users. Additionally, we filter out one-commit users and one-commit repositories to ensure robustness.

To test the statistical significance of these changes, we use an event study specification (Huntington-Klein, 2021) fit via OLS:

$$Outcome_t = \beta_1 t + \beta_2 D_{PostShock} + \beta_3 t * D_{PostShock} + \gamma_M + Intercept$$

Where the outcome is either the number of commits made by different developer groups, of first-time Rust contributors or new repos in week t, which is centered to 0 the week before the shock, the PostShock dummy equals 1 in weeks after the shock, and $\gamma_M$ denotes month-of-year fixed effects to control for natural seasonal variations in activity, for instance around Christmas and summer holidays.

We predict the following weekly outcomes in the macro models: number of commits, number of pull requests, new developers, leaving developers, and number of new and abandoned projects. Number of commits tracks the total number of commits made each week across all users, providing an aggregate measure of weekly activity in the ecosystem. The number of pull requests indicates the total number of pull requests made each week across all users and represents the amount of collaborative contributions. The number of new developers counts the number of users whose first-ever commit was recorded during the given week, indicating user inflow. The number of leaving developers identifies developers whose last recorded commit before the end of our dataset in September 2022 occurred during the given week, indicating user churn. The number of new projects reflects the count of repositories where the first-ever commit was made in a given week, tracking new ventures and ideas in the system. The number of abandoned projects measures the number of repositories in which the final commit (to September 2022) was made during the specified week, marking them as inactive thereafter.

We define six groups of developers to study potential heterogeneities in reactions to the Mozilla layoffs. We estimate the change of the activity of Mozilla users by the weekly number of commits by those users who signed a commit with an email address with a "@mozilla.com" suffix at least once during the specified period from January 1, 2019, to August 11, 2020. We detected 99 such developers.

To study differences in developers who use Rust at work or in their spare time, we follow (Riehle et al., 2014)'s method to define 8-18 developers and night owl devs. Users were classified as "8-18 developers" if at least 50% of their commits were made during weekdays (Monday-Friday) and standard working hours (8 am to 6 pm). Those who did not meet this criterion were classified as "night owls." We estimate developer time zones from commit data.

During the observation period, there are 10,215 developers who contribute primarily during weekdays from 8 AM to 6 PM, making 692,528 commits, and 15,070 developers (night owls) who contribute mainly outside of working hours, making 1,259,640 commits.

Then, non-Mozilla affiliated users are classified based on their distance from Mozilla developers in the collaboration network. To construct the network, we consider only data prior to the layoffs. We say two developers collaborate if they make contributions to the same repository, filtering out those developers who contribute less than four times to a specific repo (the median number of contributions observed). Distances in the resulting network offer a coherent measure of distance between collaborators: individuals who contribute to the same repo are direct neighbors, while developers who have a collaborator in common but do not contribute to the same repo are two steps apart, etc.

Using this network we can define four categories of developers. Mozilla developers are in their own category. All other developers are put into categories as a function of their distance in the network to Mozilla developers. Distance 1 developers are direct collaborators of Mozilla developers and distance 2 developers collaborate with direct collaborators of Mozilla developers. All developers farther away in the network, i.e. at least 3 steps away, are in the fourth group. Developers who began their activity after the shock are not included in the network. The collaboration network includes 8,870 developers responsible for 1,433,964 commits during the observation period. Of these, 1,251 developers, classified in the first category, made 338,508 commits; 3,245 developers in the second category made 580,758 commits; and 4,374 developers in the third category made 514,698 commits.

# Results

## Macro Ecosystem Outcomes

We first present a visualization of commits to all OSS Rust libraries across the Mozilla layoffs in Figure 3. We observe a break aligning with our descriptive analysis: overall activity in the ecosystem decreased relative to trend after the layoffs.

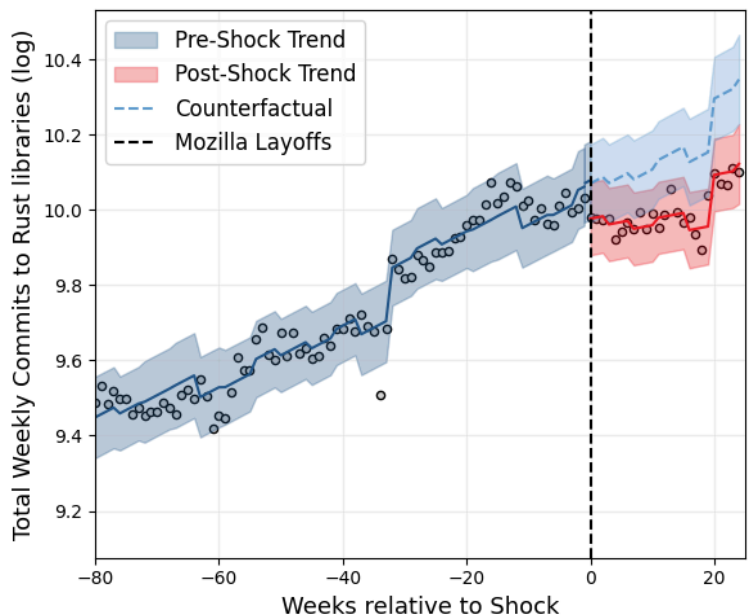

*Figure 3: Visualization of the event study model of the weekly number of commits to OSS Rust libraries. We visualize the pre- and post-layoff model fits to the data, and extrapolate the pre-shock model trend using the model to generate a no-shock counterfactual.*

|  | Commits (log) (1) | Pull Requests (log) (2) | New Devs (log) (3) | Leaving Devs (log) (4) | New Repos (log) (5) | Abandoned Repos (log) (6) |
|---|---|---|---|---|---|---|
| Time (weeks) | 0.009*** | 0.010*** | 0.006*** | 0.009*** | 0.008*** | 0.010*** |
|  | (0.000) | (0.000) | (0.001) | (0.001) | (0.000) | (0.001) |
| Shock | -0.094*** | -0.136*** | -0.094* | -0.089 | -0.098*** | -0.064 |
|  | (0.018) | (0.030) | (0.053) | (0.068) | (0.037) | (0.056) |
| Time × Shock | -0.006*** | -0.001 | -0.010*** | -0.000 | -0.009*** | 0.002 |
|  | (0.001) | (0.002) | (0.003) | (0.004) | (0.003) | (0.004) |
| # Obs. (Weeks) | 110 | 110 | 110 | 110 | 110 | 110 |
| Month-of-Year FE | X | X | X | X | X | X |
| Adjusted $R^2$ | 0.96 | 0.94 | 0.62 | 0.84 | 0.78 | 0.84 |

*Table 1: Event study models of macro-level activity in the Rust OSS ecosystem. Models include month of year fixed effects to control for seasonality. We report HAC-robust standard errors.*

Table 1 reports all six models of macro activity. We observe an immediate effect of the layoffs on the overall number of commits, pull requests and on the number of new repos created (Shock variable, models 1, 2 and 5). We observe a decrease in the trend of commits, new developers and new repos (interaction term, models 1, 3, and 5). There is no significant tendency for increased developer departure or abandonment of projects. Overall, inflow of new developers and ideas are affected, while churn and project closure rates are not.

Next we carry out a counterfactual analysis to generate more interpretable estimates of the impact of the Mozilla layoffs on various outcomes. Specifically, we ran Monte Carlo simulations using estimated model coefficients of variable trends prior to the layoffs, drawing 10,000 samples from a multivariate normal distribution of the estimated coefficients to generate a no-shock counterfactual scenario. The estimated total difference between the observed and counterfactual trajectories at week 20 was used to quantify the shock's impact, with 95% confidence intervals calculated from the simulations. We report estimated effects in percentage to facilitate interpretation.

The results of this analysis are presented in Figure 4. Here we see the asymmetry of the effect of the Mozilla layoffs between inflow and outflow more clearly. 20 weeks after the layoffs, the creation of new repos and the participation of new developers were both more than 20% lower than predicted by the counterfactual trend. Neither developer churn nor the rate of project abandonment were significantly different than predicted by the counterfactual trend.

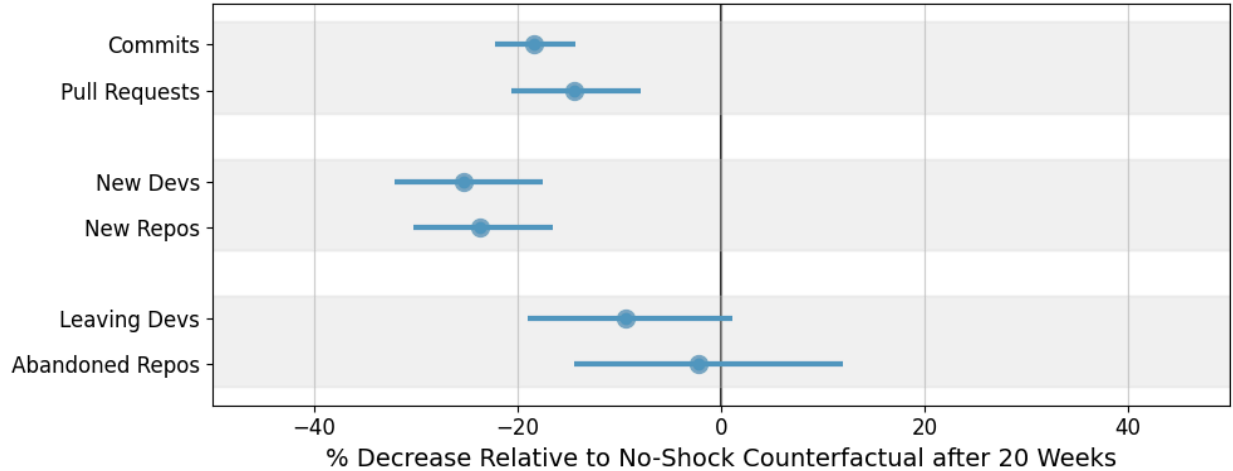

*Figure 4: Estimated changes in macro-level activity in the Rust OSS ecosystem relative to a no-shock counterfactual.*

## Developer group heterogeneities

We next move on to the analysis of various developer groups. In Table 2 we report the results of our regression models. The immediate effect of the Mozilla layoffs was significant in all six developer groups (Shock estimate, models 1-6). Unsurprisingly, commits from Mozilla developers fell by roughly 50% immediately following the layoffs. Other groups had more muted but still significant declines in the short term. Observing the changes in trends reveals important differences between the groups. Mozilla developers have a significant post-shock recovery, while contributions from developers at least two steps from them in the collaboration network shifts from a positive trend to a negative trend.

|  | Mozilla Devs (log) (1) | Collab Dist 1 (log) (2) | Collab Dist 2 (log) (3) | Collab Dist $\geq 3$ (log) (4) | 8-18 Devs (log) (5) | Night Owl Devs (log) (6) |
|---|---|---|---|---|---|---|
| Time (weeks) | 0.001 | 0.003*** | 0.006*** | 0.011*** | 0.007*** | 0.010*** |
|  | (0.001) | (0.000) | (0.000) | (0.000) | (0.000) | (0.000) |
| Shock | -0.587*** | -0.183*** | -0.174*** | -0.312*** | -0.111*** | -0.068*** |
|  | (0.102) | (0.045) | (0.031) | (0.044) | (0.038) | (0.022) |
| Time × Shock | 0.037*** | -0.003 | -0.013*** | -0.019*** | 0.004 | -0.008*** |
|  | (0.008) | (0.003) | (0.003) | (0.003) | (0.004) | (0.002) |
| # Obs. (Weeks) | 110 | 110 | 110 | 110 | 110 | 110 |
| Month-of-Year FE | X | X | X | X | X | X |
| Adjusted $R^2$ | 0.12 | 0.56 | 0.81 | 0.91 | 0.75 | 0.94 |

*Table 2. Event study models of changes in activity by various developer groups. Models include month of year fixed effects to control for seasonality. We report HAC-robust standard errors.*

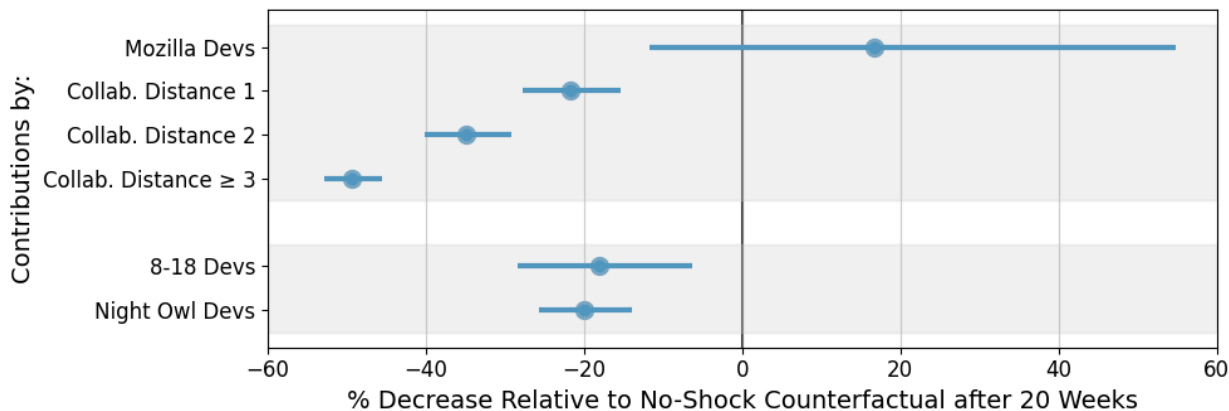

*Figure 5: Estimated changes in commits by various developer groups following the Mozilla layoffs, relative to a no-shock counterfactual.*

Simulations, carried out in the same way as for the ecosystem-wide variables, again help us to interpret the models. Within 20 weeks, the significant decrease in commits from Mozilla developers at the time of the layoffs was completely reversed. Indeed our estimate of the difference between the observed data and the counterfactual after 20 weeks is not statistically significant for Mozilla developers. For developers classified by their distance in the collaboration network from Mozilla developers, the decrease remains significant. Moreover, the decline becomes more pronounced as we move further from the core represented by Mozilla developers. We visualize the time series of direct collaborators of Mozilla developers with those at least three steps away in the network in Figure 6.

It is instructive to contrast the differing outcomes of developers classified by their distance in the collaboration network from Mozilla developers with the lack of a difference in outcomes when classified as 8–18 developers versus night owls. Surprisingly, the Mozilla layoffs appear to have been equally discouraging to developers contributing both during and outside of standard working hours.

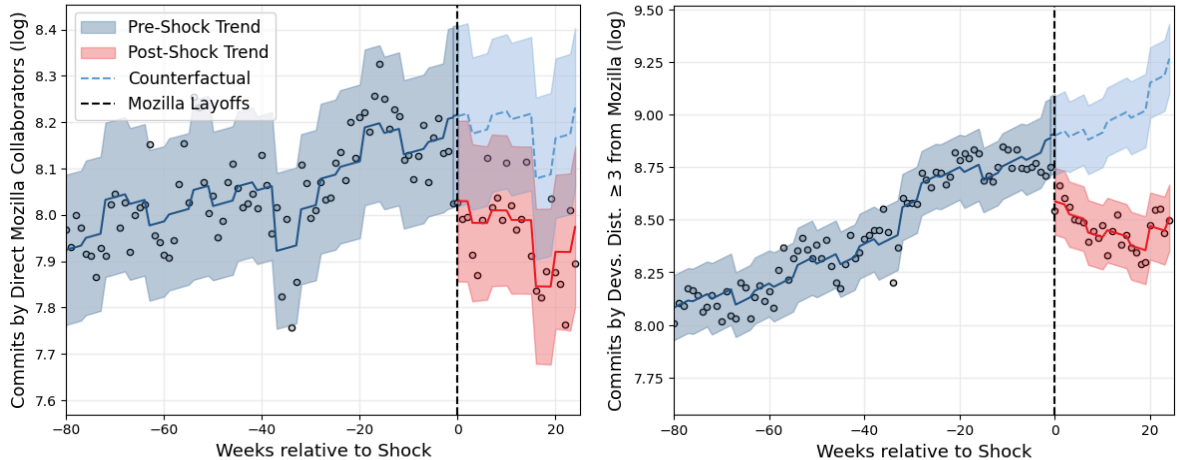

*Figure 6: Visualizing changes in activity across the shock relative to the counterfactual by direct collaborators of Mozilla developers, and those more distant in the collaboration network, respectively.*

## A toy model of an OSS ecosystem with an Anchor Sponsor

Here we develop a simple agent-based model to represent participation dynamics within an open source ecosystem including an anchor sponsor. We aim to reproduce the empirical observation that while the withdrawal of corporate sponsorship significantly depresses the inflow of new contributors, established developers—who have already accrued investment in the community—tend to remain engaged. In our framework, each developer is characterized by an intrinsic interest in participating, which is initially drawn from a standard normal distribution. This interest level is subject to enhancement via two channels: a direct sponsorship bonus when the sponsor is active and an incremental investment that accumulates with each period of continued participation.

At each discrete time period, active developers have a chance to refer new developers into the ecosystem. The referral process is stochastic, governed by a fixed referral rate, and the new entrants' initial interest is boosted by the sponsorship bonus if the sponsor is active. However, only those new developers whose combined interest (intrinsic plus any bonus) exceeds a threshold (set at zero in our model) join the community as contributors. This mechanism serves as a proxy for the higher initial credibility or resources that a sponsor can provide, lowering the entry barrier for new contributors. Conversely, once the sponsor exits at a predetermined time, this bonus is removed, making it less likely for new developers to join the system—consistent with our empirical findings.

In parallel, the model incorporates attrition through two mechanisms. First, all developers face a constant probability of random departure, capturing natural attrition. Second, the effective interest of a developer is updated each period by adding an increment proportional to their tenure reflecting their investment into the system and, while the sponsor is active, a sponsorship bonus.

Should a developer's updated interest fall below zero, they are assumed to exit the community. This dynamic ensures that developers who have invested longer in the ecosystem accumulate sufficient commitment to withstand the loss of sponsorship, thereby explaining the observed stickiness of incumbent contributors even after sponsor exit.

The model has the following parameters. We indicate the values we used to create the results below.

1. *Time Periods (100)*: The total number of time periods over which the simulation runs.
2. *Sponsor Exit Period (50)*: The specific time period at which the corporate sponsor leaves the ecosystem, removing the sponsorship bonus from that point onward.
3. *Initial Developers (5,000)*: The initial number of developers present in the system when the simulation begins.
4. *Referral Rate (0.03)*: The probability that any active developer will refer a potential new developer in each time period.
5. *Sponsorship Bonus (1)*: The additional increment added to a developer's interest when the sponsor is active. This bonus affects both the initial interest of new developers and the ongoing interest of current developers during each period.
6. *Investment Rate (0.05)*: The per-period increase in a developer's interest, reflecting growing investment or commitment over time.
7. *Natural Attrition (0.01)*: The constant probability that any developer will exit the ecosystem in a given period due to random factors.

We run the model 100 times, and report the average number of new developers and active developers in each time period in Figure 7. We observe that the model reproduces the general empirical observation that the departure of a sponsor decreases the likelihood that new contributors will join.

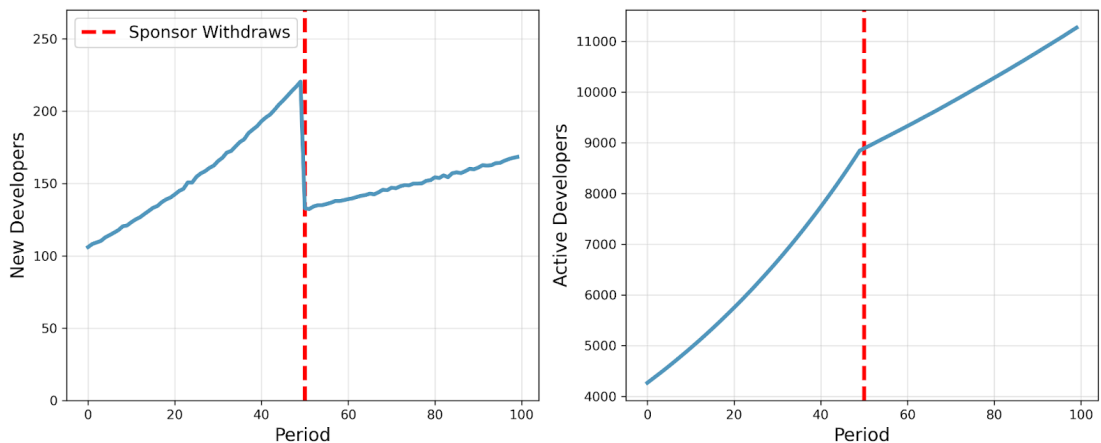

*Figure 7: Results of an agent-based model of a participation in an OSS ecosystem with an anchor sponsor. Left: number of new developers joining the ecosystem each period. Right - the total number of active developers.*

# Discussion

In this paper we studied the effect of the sudden departure of Mozilla from the Rust OSS ecosystem as a major corporate sponsor. We found a significant decrease in the number of weekly commits: 9% at the moment of the shock, and a 0.6 percentage point decrease in trend relative to the pre-shock counterfactual. This decrease goes well beyond the contributions missing from laid-off Mozilla employees. We found significant heterogeneities at both the macro ecosystem level and between different types of users. While there was a sharp decrease in the rate of fire-time contributors and new projects started, there was a relatively small change in the rate of departures from the system. Grouping developers by their distance to Mozilla developers in the collaboration network, we found that developers closer to Mozilla had smaller decreases in activity than those further away. Contributions by developers far from Mozilla remained below trend even half a year after the shock. We observed no heterogeneity between developers who mainly contribute during working hours and those who contribute nights and weekends.

Together, these results suggest an asymmetric influence of anchor sponsorship on developer behavior: an anchor sponsor's presence encourages entry and innovation by reducing perceived risks through signaling. Developers that are already established and invested in an ecosystem, are less affected by the anchor sponsor's presence or departure. A simple agent-based model that combines developer investment over time with a bonus from the presence of an anchor sponsor reproduces these qualitative observations.

We highlight three contributions of our work. First, we extend understanding of OSS ecosystem sustainability and evolution in the context of growing anchor sponsorship by firms. Our findings show that an anchor signals quality and stability that attracts newcomers, but can leave the system vulnerable to sudden changes in support - similarly to a region's vulnerability to the collapse of an anchor firm (Ornston and Camargo, 2024). Second, we provide insights into developer motivations relative to significant firm involvement, as core participants often remain engaged despite sponsor exit, while distant or new contributors hesitate or leave. Finally, our findings suggest that the analogy between OSS ecosystems and clusters generates productive theoretical insights.

The results suggest that anchor sponsors shape developer behavior through knowledge ties and reputational effects (Agrawal and Cockburn, 2003; Feldman, 2003; Pashigian and Gould, 1998). Direct collaborators tend to stay because they have stronger social and technical capital in the ecosystem (Ranft and Lord, 2000), but more distant developers move away or refrain from joining (Qiu et al., 2019a; Smirnova et al., 2022). We also note that sponsor withdrawal impact is more pronounced in lower inflows and continued slowdowns on the ecosystem level compared to findings of previous studies working on the project level (Zhang et al., 2018; Zhou and Mockus, 2014). These patterns reflect a combination of intrinsic and extrinsic motivations, where firm involvement attracts new participants (Alexy, 2009; Lerner and Tirole, 2002) but also creates a risk of decline when that involvement ends.

Several limitations of our work suggest fruitful lines of research to be investigated. Our focus on a single ecosystem with a single anchor limits generalization. The major drawback to our macro-study of the Rust ecosystem is that we are only observing one specific event. Future work might compare multiple ecosystems with several sponsors to assess how diversification buffers shocks. It could also examine how volunteers, hobbyists, and corporate employees differ in their responses to sponsor signals, or use inter-project dependencies to define alternative notions of proximity to Mozilla or other anchor sponsors (Decan et al., 2019; Schueller and Wachs, 2024).

Indeed, it is difficult to extrapolate what would have happened had Mozilla abandoned Rust in 2017, or if Mozilla had had an even larger footprint at the time of the layoffs. Perhaps Rust itself was already attractive enough to remain sustainable: by late 2020, Rust was widely used and there was a large enough labor market to absorb many of these layoffs. By February 2021, Amazon, Huawei, Google, Microsoft and Mozilla announced the creation of the Rust Foundation to manage the language, its trademarks and domains. In this way, efforts were made to restore indirect firm support and to signal Rust's quality. Overall, while the Rust experience highlights some of the strengths and drawbacks of the increasingly common phenomenon of anchor sponsorship in OSS ecosystems, more work is needed to understand this emerging model of collaborative work.